\documentclass[,final] {aipproc}
\usepackage{amssymb}
\usepackage{amsfonts}
\usepackage{amsmath}

\layoutstyle{8x11single}
\begin{document}

\title{From Elastic Continua to Space-time}


\author{A.Tartaglia}{
  address={Dipartimento di Fisica, Politecnico, Corso Duca degli Abruzzi 24, 10129 Torino and INFN, Sezione di Torino, Via Pietro Giuria 1, 10126 Torino}
}

\author{N.Radicella}{
  address={Dipartimento di Fisica, Politecnico, Corso Duca degli Abruzzi 24, 10129 Torino and INFN, Sezione di Torino, Via Pietro Giuria 1, 10126 Torino}
}

\begin{abstract}
Since the early days of the theory of electromagnetism and of gravity the idea of space, then space-time, as a sort of physical continuum hovered the scientific community. Actually general relativity shows the strong similarity that exists between the geometrical properties of space-time and the ones of a strained elastic continuum. The bridge between geometry and the elastic potential, as well in three as in three plus one dimensions, is the strain tensor, read as the non-trivial part of the metric tensor. On the basis of this remark and exploiting appropriate multidimensional embeddings, it is possible to build a full theory of space-time, allowing to account for the accelerated expansion of the universe. How this can be obtained is the content of the paper. The theory fits the cosmic accelerated expansion data from type Ia supernovae better than the $\Lambda CDM$ model. 
\end{abstract}

\maketitle

\section{Introduction}
The description contemporary physics gives of the universe is essentially based on a dualistic approach which is perfectly expressed by the Einstein equations: 	
\begin{equation}
G_{\mu\nu}=k T_{\mu\nu}
\end{equation}

As we well know, the right hand side contains matter/energy (whatever this means) and the left hand side is 'geometry': marble on the left, wood on the right, according to a definition attributed to Einstein himself. However 'geometry' is rather abstract: it actually has to be the geometry of something and this something is space-time. The queer questions do not end here: what is space-time? It cannot simply be a mathematical artifact, since it does interact with matter/energy producing physical effects. It has to be some physical entity though different in properties from what we know as matter. If it is so, space-time is indeed provided with its own physical properties which in turn tradition-ally appear in the action integral via the second order derivatives of the elements of the metric tensor and the minimal coupling to the matter fields in the energy-momentum tensor. 
However space-time has manifest similarities with material continua, even though the latter are a macroscopic approximated description of a microscopically discontinuous situation, which is not the case for space-time, at least as far as we do not try and quantize it. Quoting Einstein, who actually was speaking of space rather than space-time, '{\it\dots  according to the general theory of relativity space is endowed with physical qualities; in this sense, therefore, there exists an ether\dots But this ether may not be thought of as endowed with the quality characteristic of ponderable media, as consisting of parts which may be tracked through time\dots }'\cite{einstein}. What we are doing in the present paper is to take this similarity with material continua seriously, while keeping all requirements typical of General Relativity (GR) as local Lorentz invariance and general covariance. Our four-dimensional continuum will be a Rie-mannian manifold, whose intrinsic curvature will be thought as been due to a deformation from a reference flat state. Since this is the case for ordinary material continua we shall also introduce the concept of texture defects as sources of spontaneous strain in the bulk, which in our case means curvature even in the absence of matter. Matter/energy will then be treated, as it usually is, as an additional source of curvature (then strain). This approach will prove to be appealing and effective in describing many properties we know to be present in the actual universe.

\section{Embedding and basic concepts}
\begin{figure}[htb]
\includegraphics[width=13cm]{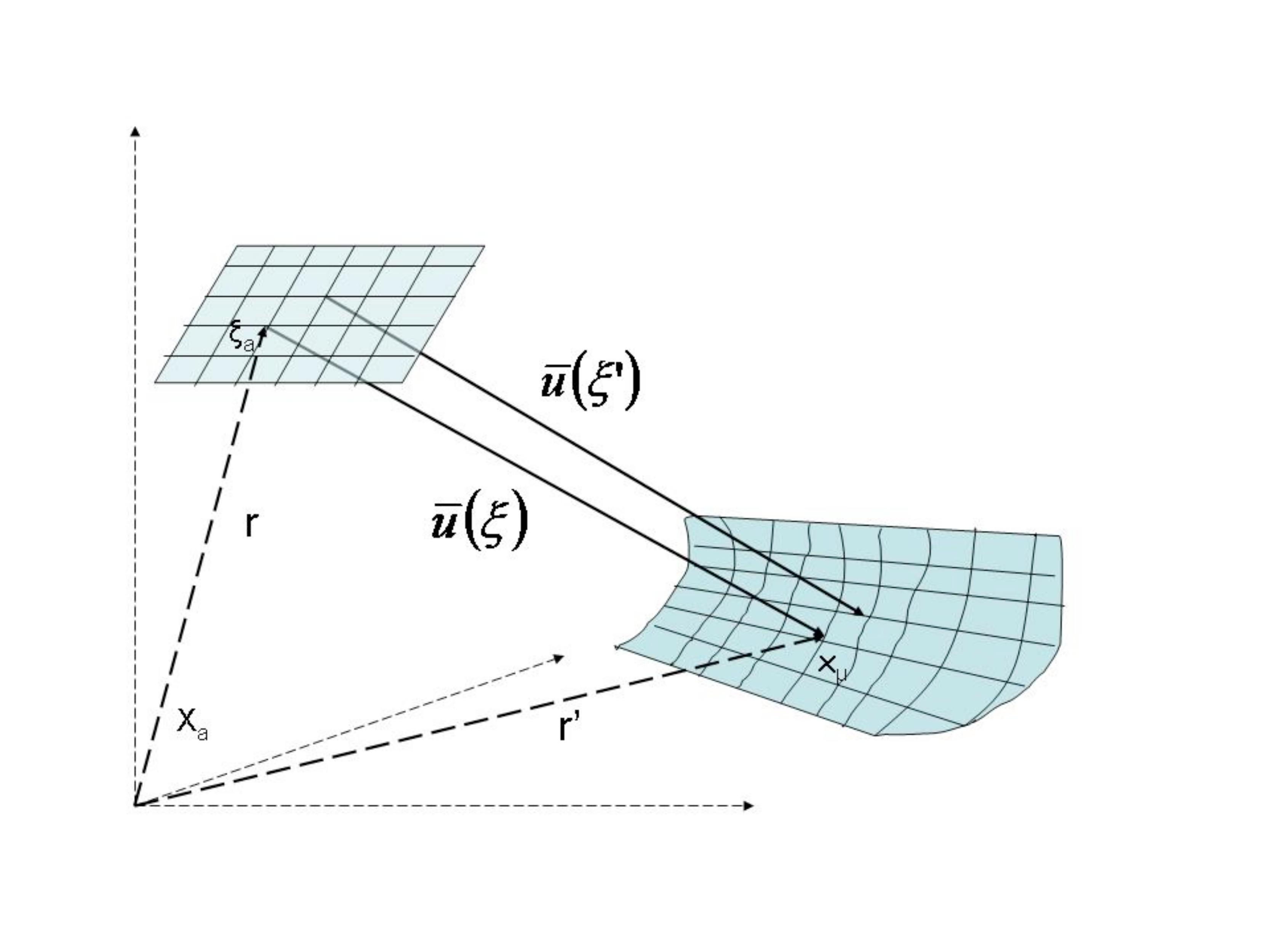}
\caption{Two $N$-dimensional manifolds are shown embedded in an $(N+1)$-dimensional one. The embedding manifold is flat, as well as one of the N-dimensional manifolds; the other, represented on the right, is instead intrinsically curved. The $(N+1)$ dimensional r vectors point to events of the embedding which lie on the two lower dimensional manifolds. The $u$ displacement vectors establish a one-to-one correspondence between the two sub-manifolds. $X$'s are the coordinates of the flat embedding, $\xi$'s are the ones of the flat sub-manifold, $x$'s the ones of the curved sub-manifold }
\label{fig00}
\end{figure}

We start considering an $(N+1)$-dimensional embedding manifold. This manifold is assumed to be flat, with Lorentzian signature. Let us establish a coordinate  ( $X$'s) on it. In this embedding we consider two N-dimensional sub-manifolds. One, which we shall call the reference manifold, is assumed to be flat also. The coordinates on it are the $\xi$'s. The other sub-manifold, which will be called the natural one \cite{landau, eshelby}, is assumed to be intrinsically curved and endowed with the x coordinates. Of course the choice of the coordinates in the various manifolds is entirely free. The dimensional reduction in the sub-manifolds from $N+1$ to $N$ dimensions is performed  by means of appropriate constraint equations, one per sub-manifold:
\begin{equation}
f\left(X_1,X_2,\dots,X_{N+1}\right)=0
\end{equation}
The constraining functions are assumed to be continuous and differentiable as needed by all further developments                              
We may localize events on the main manifold by means of $(N+1)$-radii from the origin of the $X$ coordinates system. If the two sub-manifolds are smooth we may establish a one to one correspondence between them connecting pairs of events on each by means of appropriate $(N+1)$-dimensional displacement vectors; let us call them u's. Since the embedding manifold is flat the u vectors may easily be parallel-transported around and are everywhere well de-fined and well behaved. The u's are a vectorial displacement field based on the reference manifold and landing on the natural one, so that it may be described either in terms of the $\xi$'s or of the $x$'s. Typically it will be:
\begin{equation}
u(X)=r'(X)-r(X)
\end{equation}
The situation is represented in figure \ref{fig00}.

The differential change in the u's 
$$
\frac{\partial u^a}{\partial X^b}=\frac{\partial u^a}{\partial \xi^\mu}\frac{\partial \xi^\mu}{\partial X^b}=\frac{\partial u^a}{\partial x^\mu}\frac{\partial x^\mu}{\partial X^b}, \quad\quad a=1,\dots N+1, \quad\mu=1, \dots, N,
$$
expresses the presence of strain in the curved manifold.  
The next step is to write down the metric properties of the manifolds. Considering two close-by events in the reference manifold we may write the squared distance (interval) between them as
\begin{equation}\label{flat}
dl^2=\eta_{ab} dX^a dX^b=\eta_{ab}\frac{\partial X^a}{\partial \xi^\mu}\frac{\partial X^b}{\partial \xi^\nu} d\xi^\mu d\xi^\nu=\eta_{\mu\nu} d\xi^\mu d\xi^\nu
\end{equation}
Since the main, as well as the reference, manifold is flat, the $\eta$'s are the components of the Minkowski metric tensor. We may then consider the distance between the two events in the natural manifold, corresponding to the pair in the reference. It will be
\begin{equation}\label{curve}
dl'^2=\eta_{ab} dX'^a dX'^b=\eta_{ab}\frac{\partial X'^a}{\partial x^\mu}\frac{\partial X'^b}{\partial x^\nu} dx^\mu dx^\nu=g_{\mu\nu} dx^\mu dx^\nu
\end{equation}
Since the natural manifold is curved the metric tensor is no more Minkowskian. Combining eq.s (\ref{flat}) and (\ref{curve}) and ex-pressing everything by means of the coordinates of the natural manifold we have
\begin{equation}
g_{\mu\nu}=\left(\eta_{\mu\nu}+2\varepsilon_{\mu\nu}\right)
\end{equation}
being  $\varepsilon_{\mu\nu}$ the strain tensor, depending on the differential changes of the displacement field. The explicit formula for the elements of the strain tensor is
\begin{equation}
\varepsilon_{\mu\nu}=\eta_{a\nu}\frac{\partial u^a}{\partial x^\mu}+\eta_{b\mu}\frac{\partial u^b}{\partial x^\nu}+\eta_{ab}\frac{\partial u^a}{\partial x^\mu}\frac{\partial u^b}{\partial x^\nu}
\end{equation}
The formal treatment we have done so far can be interpreted in physical terms thinking of an initial situation represented by the flat unstrained reference manifold; then some external action (in three dimensions we would speak of forces) is applied, producing a deformation of the manifold and leading to the final situation, represented by the natural manifold. If the change of the metric tensor could be expressed in the form
$$
g_{\mu\nu}=\eta_{\alpha\beta}\frac{\partial \xi^a}{\partial x^\mu}\frac{\partial \xi^\beta}{\partial x^\nu}
$$
the new metric would be diffeomorphic to the initial flat one. Actually De Saint VenantÕs integrability condition 
\begin{equation}
R^{\alpha}_{\beta\mu\nu}=0
\end{equation}
would be satisfied, being $R^{\alpha}_{\beta\mu\nu}$ the curvature tensor. In practice this is the case of pure elastic deformations where no intrinsic curvature is present: the strain cannot be detected from within the deformed manifold.

\section{Defects in the Manifold}
We are interested in intrinsic curvature, so we should look for situations more general than pure elasticity. Actually in material elastic continua defects can be present. The idea of a defect in the texture of a continuum has been analyzed since the pioneering work of Vito Volterra in 1904 \cite{volterra}, in connection with the study of plastic deforma-tions. In our case the idea of defect is visually expressed in fig. \ref{fig2}.
\begin{figure}[htb]
\includegraphics[width=8cm]{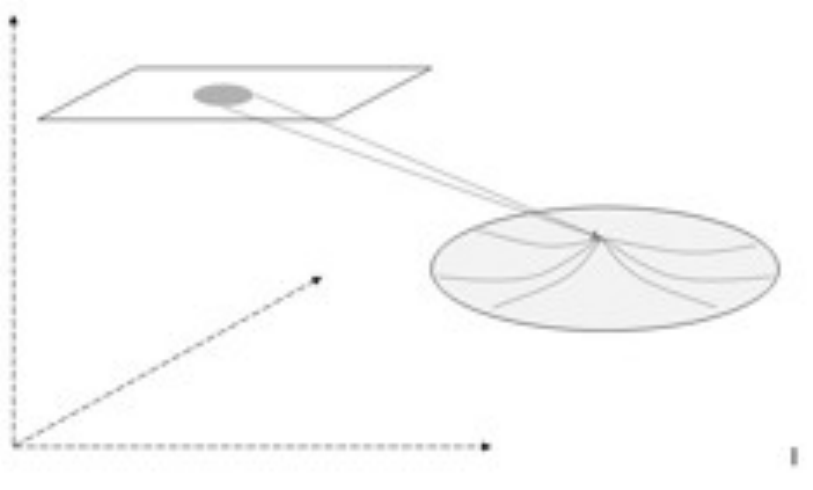}
\caption{A defect in the right hand side manifold is a point corresponding to a whole domain of the reference flat manifold. Actually the defect is represented by a set of points in the natural manifold with a lower dimensionality than the corresponding set in the reference manifold. A defect is generally associated with spontaneous strain and curvature. }
\label{fig2}
\end{figure}

When a set of points in the natural manifold corresponds to a higher dimensional set in the reference manifold we have a defect. We may think the defect as the result of a cut and successive sewing of the borders in the original manifold, that, by so doing, is transformed into the natural manifold. This process would in general produce intrinsic curvature i.e. a spontaneous strain state. Generally speaking the transformation from the coordinates of the reference manifold to the coordinates of the natural manifold may be written as a one-form
\begin{equation}\label{form}
dx^\mu=\Phi^\mu_\alpha d\xi^\alpha.
\end{equation}
In the case of a pure elastic deformation the transformation is a diffeomorphism and (\ref{form}) is a closed one-form. When a defect is present (\ref{form}) is a general one-form, which means that choosing a closed path encircling the defect it will be
$$
\oint \Phi^\mu_\alpha d\xi^\alpha\neq 0.
$$
Besides introducing a spontaneous strain state, the presence of a defect also fixes the global symmetry of the manifold.
 
 \section{A Lagrangian for the 'Elastic' Space-time}
 Once we have decided to describe space-time as a material continuum in the way exposed in the previous sec-tion, if we wish to find the equilibrium configuration we need to write down an appropriate Lagrangian accounting for the properties we want to include.  The starting point is the known Hilbert-Einstein action used for GR:
 \begin{equation}\label{eh}
 \int R\sqrt{-g} d^4x
 \end{equation}
 $R$ is the scalar curvature and g is the determinant of the metric tensor used to define the invariant integration four-volume.
An important difference between our theory and the classical elasticity theory in three dimensions is that in our case all variables are in the manifold, whereas in the elasticity theory time is an evolution parameter which is not part of the manifold. Keeping this in mind we notice that the integrand in (\ref{eh}) may be interpreted as a ÒkineticÓ term since it contains the derivatives of the elements of the metric with respect to the coordinates; actually they are second order derivatives, which, appearing linearly, can be reduced to first order ones by means of an integration by parts. According to the typical structure of a classical Lagrangian we should add a potential term too, accounting for the deformations of the basic manifold.
In the classical theory of elasticity the typical elastic potential energy would be written
 \begin{equation}\label{potential}
 \frac{1}{2}\sigma_{\mu\nu}\varepsilon^{\mu\nu}.
 \end{equation}
 The  $\sigma$Õs now are the components of the stress tensor and the stress tensor in turn is a function of the strain tensor. The type of relation between stress and strain depends on the properties of the material we are considering and we have a priori no idea of what it could be for space-time. However in three dimensions the low strain approximation is of course the linear one, good for most applications. Let us then assume that the linear dependence is good for space-time too, so we assume and write:
 \begin{equation}\label{stress}
 \sigma_{\mu\nu}=C_{\mu\nu\alpha\beta} \varepsilon^{\alpha\beta}.
 \end{equation}
 The rank 4 tensor C is the elastic modulus tensor. Both $\varepsilon$ and $\sigma$ are symmetric tensors, so considering the general symmetries if the medium (for us: the space-time) is homogeneous and isotropic all elements of C depend on two parameters only, $\lambda$ and $\mu$ , called the Lam\'e coefficients. We write:
 \begin{equation}
C_{\alpha \beta \mu\nu }=\lambda \eta _{\alpha \beta }\eta _{\mu
\nu }+\mu \left( \eta _{\alpha \mu }\eta _{\beta \nu }+\eta
_{\alpha \nu }\eta _{\beta \mu }\right)  \label{moduli}
\end{equation}
Since all quantities are tensors and the equations express local relations everything is referred to the tangent space and the ?Õs account for the Lorentzian signature on it.
The explicit form of  (\ref{stress}) is now
\begin{equation}
\sigma^{\mu\nu}=\lambda \varepsilon \eta^{\mu\nu}+2\mu \varepsilon^{\mu\nu}
\end{equation}
being $\varepsilon=\varepsilon^\alpha_\alpha$ the trace of the strain tensor.
Summing up, using eq. (\ref{potential}) and introducing matter,  we are finally able to write down a complete action integral for space-time plus external sources, i.e. matter. This is 
\begin{equation}\label{action}
S=\int\left(R+\frac{1}{2}\lambda \varepsilon^2+2\mu \varepsilon_{\mu\nu}\varepsilon^{\mu\nu}+\kappa L_{matter}\right) \sqrt{-g} d^4 x
\end{equation}
$L_{matter}$ is the usual matter Lagrangian minimally coupled to geometry via the full metric tensor; the coupling constant is
$$
\kappa=16\pi\frac{G}{c^2}
$$
Once the Lagrangian has been read out of (\ref{action}) it is possible to obtain a generalized Einstein equation in the form
\begin{equation}\label{EEE}
G_{\mu\nu}=T_{e\mu\nu}+\kappa T_{\mu\nu}
\end{equation}
$G_{\mu\nu}$ is the Einstein tensor, $T_{e\mu\nu}$ is the effective energy/momentum tensor obtained from the 'elastic' potential energy term in the Lagrangian, the rest is the usual matter source. 

\section{A Robertson-Walker Universe}          
\begin{figure}[htb]
\includegraphics[width=10cm]{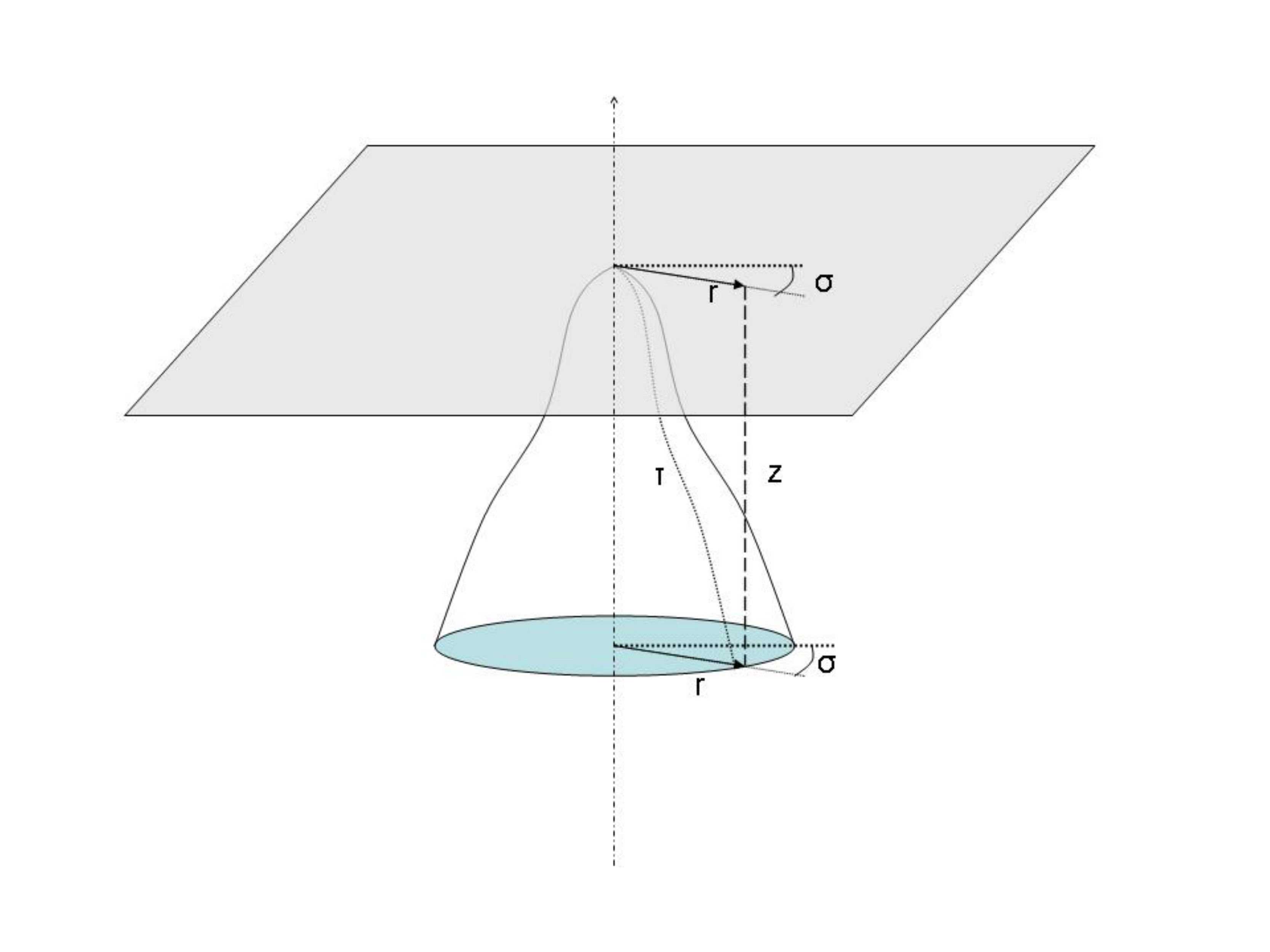}
\caption{An example of a Robertson-Walker symmetric bidimensional manifold embedded in a three-dimensional flat space-time, is shown. The light gray surface is the flat reference manifold. The r coordinate is time-like, ? is the transverse, i.e. space-like, line element; both $r$ and $\sigma$ are the same for the embedding and the reference manifold. The bell-shaped figure represents the natural manifold; the ? variable is the same as the one of the reference manifold, but the ÒradialÓ coordinate is now the cosmic time $\tau$. The cosmic time is a function of $r$; the functional form depends on the type of defect leading from the reference to the natural manifold. The present example is valid for a closed space.}
\label{emb}
\end{figure}

People usually think that the universe at large is endowed with a global symmetry allowing for a global foliation with homogeneous and isotropic space sheets plus a cosmic time: this is the typical Robertson-Walker (RW) symmetry. Why is that symmetry there? Actually the presence of matter in the form of dust plus radiation does not guarantee a RW symmetry. Here, in our approach, comes the role of a space-time defect, which we call cosmic defect (CD). The defect defines and fixes the global symmetry on the natural manifold; the location of the defect coincides with the initial singularity of a standard GR cosmology. The situation is most effectively represented in fig.\ref{emb} which envisages bidimensional manifolds in a three-dimensional embedding.
One may think of obtaining the curved natural manifold from the flat reference by means of an appropriate cut and sew process. The shape of the cut determines the final configuration of the natural manifold. The space coordinates are represented by the ? in the figure, whilst the radial variables are times. The cosmic time ? of the natural manifold is a function of the r of the reference manifold, through the z coordinate of the embedding manifold. Here the Robertson-Walker symmetry is a reflex of an axial symmetry in the embedding. 
We can now write down corresponding line elements in the reference and natural manifolds. For the flat Minkowski reference one has 
\begin{equation}
ds^{2}=dr^{2}+dz^{2}-r^{2}d\sigma ^{2}=dr^{2}-r^{2}d\sigma ^{2}
\label{lpiatto}.
\end{equation}
In the natural manifold it is
\begin{equation}
ds^{\prime 2}=d\tau ^{2}-r^{2}d\sigma ^{2}=\left( 1+f^{\prime 2}\right).
dr^{2}-r^{2}d\sigma ^{2}  \label{lcurvo}
\end{equation}
Use has been made of the fact that $d\tau^2=dr^2+dz^2$ and $z=f(r)$; $f'$ id the $r$ derivative of the function $f$. \\
The difference between (\ref{lcurvo}) and (\ref{lpiatto}) allows us to directly read out the relevant elements of the strain tensor, which indeed reduce to one term only:
\begin{equation}\label{e00}
\varepsilon _{00}=\frac{f^{\prime 2}}{2\left( 1+f^{\prime 2}\right) }.
\end{equation}
Of course it is better to express everything in terms of coordinates on the natural manifold; then, in order to use the same notation as in RW universes, we write $r=a(\tau)$ and $\dot{a}=da/dt$. Finally (\ref{e00}) becomes
\begin{equation}
\varepsilon _{00}=\frac{1-\dot{a}^{2}}{2}.
\end{equation}                   
The example we have worked out so far corresponds to a closed space universe. We can see this from fig.\ref{emb} where a space section of the natural manifold at constant cosmic time is represented by a circle. We have used this example because it has a simple graphical representation, however, as it is known, there are good reasons to think that space is actually flat and infinite. Though this situation does not admit a simple graphical description it is easy to treat it the same way as in the above example \cite{tarta}. The result which is obtained for the strain tensor has non-zero elements for the space-space components along the diagonal only. The explicit expression is     
\begin{equation}\label{eii}
\varepsilon _{ii}=\frac{1-a^{2}}{2}.
\end{equation}      
which will be used in the following for our cosmological applications; i is any space index.
Once the global symmetry has been fixed by the presence of the CD and (\ref{eii}) has been obtained, we can introduce both the symmetry and the latter result into the action integral (\ref{action}), which is recast in the form
\begin{equation}\label{cosmoaction}
S=\int\left[-6(a\ddot{a}+\dot{a}^2)a+\frac{9}{8} B \frac{(1-a^2)^2}{a}+\kappa a^3 L_{m}\right] d\tau.
\end{equation}
The RW symmetry has combined the two Lam\'e coefficients into the only bulk modulus of space-time
\begin{equation}\label{bulk}
B=\lambda+\frac{2}{3}\mu>0
\end{equation}
The positivity constraint has been drawn by analogy from the physical meaning of $B$ in three dimensions (an object shrinks when compressed).
From now on the process continues as usual, applying a variational principle to (\ref{cosmoaction}). Of course we need specify-ing the matter Lagrangian Lm. Let us suppose that matter is made of many different components, each with its own equation of state. It is useful, for the use in the next section, to write down explicitly the squared Hubble parameter deducible from (\ref{cosmoaction}):
\begin{equation}
\frac{\dot{a}^2}{a^2}=\frac{3}{16}B\frac{\left( 1-a^{2}\right) ^{2}}{a^4}+\kappa
\sum\limits_{i}^{n}\rho _{i0}\frac{a_{0}^{3\left( 1+w_{i}\right) }}{a^{3\left(
1+w_{i}\right) }} . \label{energia}
\end{equation}
In (\ref{energia}) H is the Hubble parameter, n is the number of matter components, $w_i$ is the state parameter of the ith component\footnote{From $p_i=w_i \rho_i$ being $p_i$ the pressure}; the 0 label refers to present day values.      

\section{The Luminosity of type IA supernovae}
A typical bench mark for cosmological theories is the dependence of the luminosity of type Ia supernovae (SnIa) on the redshift parameter z. As it is well known it is precisely the measured luminosity of SnIaÕs that revealed at the end of 1998 the accelerated expansion f the universe \cite{riess98, perlmutter99}. A typical Friedman-Robertson-Walker universe does not fit the luminosity experimental data well. Actually the best fitting model is the so called $\Lambda$-Cold-Dark-Matter ($\Lambda CDM$) theory, which implies the presence in the universe of a dark energy more than 20 times bigger than the energy content of visible matter, plus dark matter in an amount of the order of 9 times the visible matter.
In order to deduce the luminosity dependence on z of the SnIa's, we use the luminosity distance D, written as
\begin{equation}
D=m-M=25+5\log_{10} \left( \left( 1+z\right) \int_{0}^{z}\frac{dz^{\prime }}{
H(z^{\prime })}\right)   \label{modulus}
\end{equation}
Here $m$ is the observed magnitude of the source, $M$ is its absolute magnitude; the formula is correct if distances are measured in Mpc. 
Since the farthest observed supernova is at a z not bigger than 2, we consider a dust filled universe so that $w=0$, and any other contribution, including radiation, is negligible. Under this hypothesis we introduce (\ref{energia}) into (\ref{modulus})  and use $a_0$, $\rho_0$ and $B$ as optimization parameters for a best fit of the data. The fitting curve, together with the experimental values from 307 supernovae of the Supernova Cosmology Project union survey \cite{kowalski08}, is shown on fig.\ref{fig2}.
\begin{figure}[htb]
\includegraphics[width=10cm]{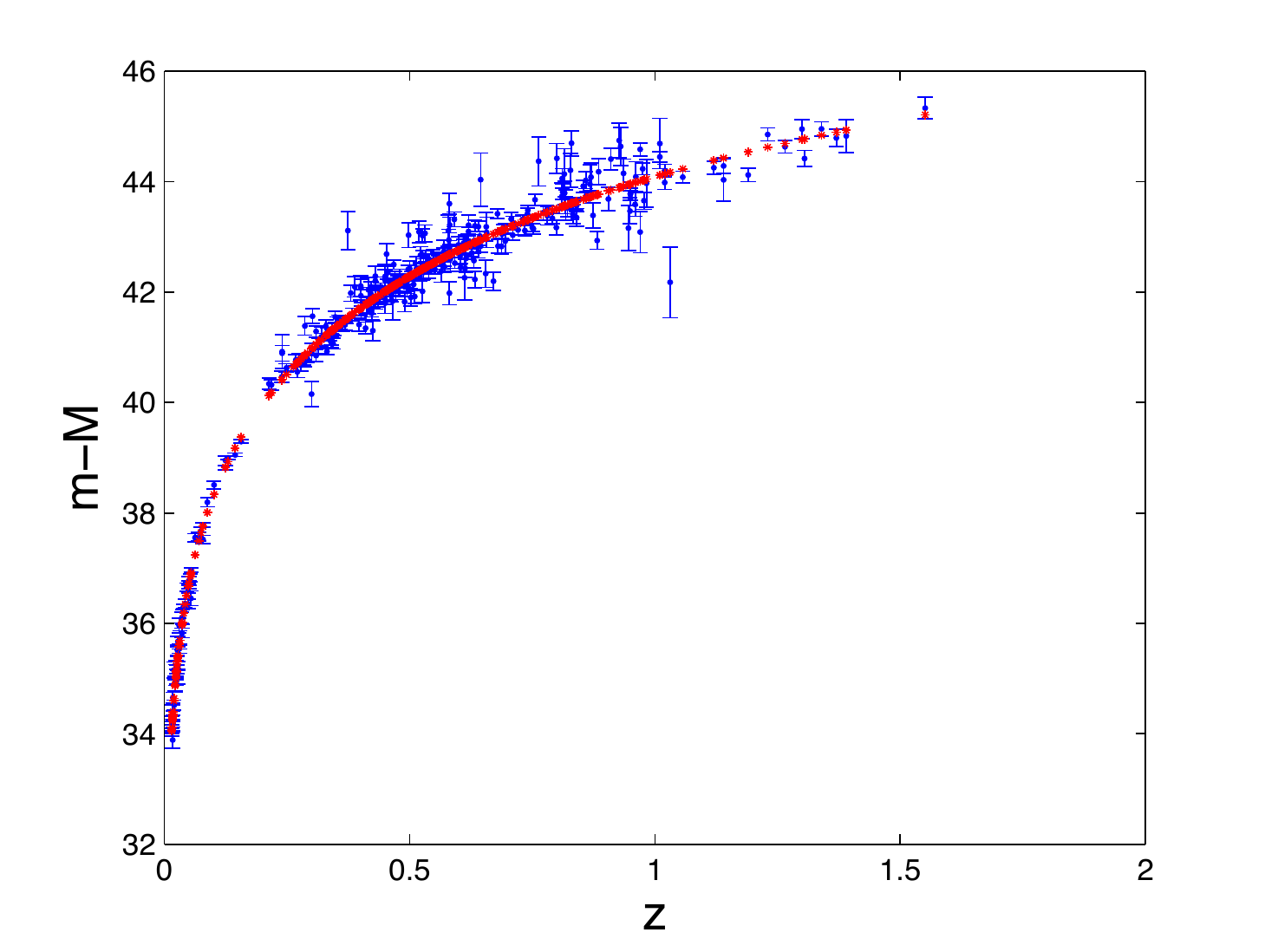}
\caption{Best fit of the luminosity distance data of SnIa's obtained using the CD theory with three optimization parameters. Dots are given by the theory; the experimental data are shown with error bars on the luminosity distance; the error on z is undetectable at the scale of the graph. 307 supernovae have been used. }
\label{fig2}
\end{figure}

The reduced $\chi^2$ of the fit is 1.017. The reduced $\chi^2$ applying the ?CDM theory to the same data in similar conditions is 1.019, so that we may claim that CD proves to be slightly better than  $\Lambda CDM$ for this purpose.
Together with the fit in fig.\ref{fig2} we have the optimization values of the parameters, which, expressed in terms of physically relevant quantities, are
\begin{equation}
\left\{ 
\begin{array}{c}
\rho_0\sim 3.4\times 10^{-27} \quad kg/m^3\\
H_0=(64\pm 35) \quad km /(s\times Mpc)\\
B=(3\pm2) \times 10^{-7} Mpc^{-2}=(3\pm2)\times 10^{-52} m^{-2}
\end{array}
\right.   \label{ottimi}
\end{equation}
The values in (\ref{ottimi}) are compatible with the currently accepted ones, without need for dark energy and dark mat-ter. The uncertainties are indeed very huge, due to the corresponding uncertainties of the experimental data. The matter density is evaluated with an almost 100\% uncertainty.

\section{Conclusions}
We have shown that the idea of a space-time endowed with physical properties and behaviours typical of mate-rial continua is indeed viable and performs well at least in the classical test of reproducing the luminosity distance curve of typa Ia supernovae. The theory preserves the general features of GR; it is a metric theory with general co-variance, tangent flat Minkowskian space, and coupling to matter via geometry. The origin of the global symmetry of the universe is ascribed to the presence of a Cosmic Defect corresponding to the origin of cosmic time. The defect is described as typical defects are in the ordinary material continua; it corresponds to a singularity in the vector displacement field leading from a flat Lorentzian manifold to the actual RW universe. In practice the actual metric tensor of our universe is described as being the sum of a constant symmetric tensor plus twice the strain tensor of space-time induced by the presence of the Cosmic Defect. The numerical values obtained for the 'elastic' parameters of space-time by fitting the luminosity distance data of supernovae prove to be compatible with the constraints posed by the behaviour of matter on a local scale. Indeed the 'elastic' parameters, because of the RW symmetry reduce to the only bulk modulus of space-time B and the order of magnitude of B is such that its effects show up only at the cosmic scale. Furthermore the Newtonian and even GR local limit of the theory is ensured by the fact that space-time is represented by a Riemannian manifold admitting everywhere (except at the defect) a tangent flat space with Lorentzian signature.

\end{document}